\RequirePackage{fix-cm}
\documentclass[twocolumn,epjc3]{svjour3}  
\smartqed  
\RequirePackage{graphicx}
\RequirePackage{verbatim}
\usepackage{lineno}
\usepackage{hyperref}
\begin{document}
\title{Search for high energy 5.5 MeV solar axions with the complete Borexino dataset}




 \author{
 D.~Basilico\thanksref{Milano}
 \and
 G.~Bellini\thanksref{Milano}
 \and
 J.~Benziger\thanksref{PrincetonChemEng}
 \and
 R.~Biondi\thanksref{MPIH,LNGS}
 \and
 B.~Caccianiga\thanksref{Milano}
 \and
 F.~Calaprice\thanksref{Princeton}
 \and
 A.~Caminata\thanksref{Genova}
 \and
 A.~Chepurnov\thanksref{Lomonosov}
 \and
 D.~D'Angelo\thanksref{Milano}
 \and
 A.~Derbin\thanksref{Peters,Kurchatov}
 \and
 A.~Di Giacinto\thanksref{LNGS}
 \and
 V.~Di Marcello\thanksref{LNGS}
 \and
 X.F.~Ding\thanksref{IHEP,Princeton}
 \and
 A.~Di Ludovico\thanksref{LNGSG,Princeton} 
 \and
 L.~Di Noto\thanksref{Genova}
 \and
 I.~Drachnev\thanksref{Peters}
 \and
 D.~Franco\thanksref{APC}
 \and
 C.~Galbiati\thanksref{Princeton,GSSI}
 \and
 C.~Ghiano\thanksref{LNGS}
 \and
 M.~Giammarchi\thanksref{Milano}
 \and
 A.~Goretti\thanksref{LNGSG,Princeton} 
 \and
 M.~Gromov\thanksref{Lomonosov,Dubna}
 \and
 D.~Guffanti\thanksref{Bicocca,Mainz}
 \and
 Aldo~Ianni\thanksref{LNGS}
 \and
 Andrea~Ianni\thanksref{Princeton}
 \and
 A.~Jany\thanksref{Krakow}
 \and
 V.~Kobychev\thanksref{Kiev}
 \and
 G.~Korga\thanksref{London,Atomki}
 \and
 S.~Kumaran\thanksref{CALI,Juelich,RWTH}
 \and
 M.~Laubenstein\thanksref{LNGS}
 \and
 E.~Litvinovich\thanksref{Kurchatov,Kurchatovb}
 \and
 P.~Lombardi\thanksref{Milano}
 \and
 I.~Lomskaya\thanksref{Peters}
 \and
 L.~Ludhova\thanksref{GSIMAINZ,Juelich,RWTH}
 \and
 I.~Machulin\thanksref{Kurchatov,Kurchatovb}
 \and
 J.~Martyn\thanksref{Mainz}
 \and
 E.~Meroni\thanksref{Milano}
 \and
 L.~Miramonti\thanksref{Milano}
 \and
 M.~Misiaszek\thanksref{Krakow}
 \and
 V.~Muratova\thanksref{Peters}
 \and
 R.~Nugmanov\thanksref{Kurchatov,Kurchatovb}
 \and
 L.~Oberauer\thanksref{Munchen}
 \and
 V.~Orekhov\thanksref{Mainz}
 \and
 F.~Ortica\thanksref{Perugia}
 \and
 M.~Pallavicini\thanksref{Genova}
 \and
 L.~Pelicci\thanksref{MI,Juelich,RWTH}
 \and
 \"O.~Penek\thanksref{BOS,Juelich}
 \and
 L.~Pietrofaccia\thanksref{LNGSG,Princeton}
 \and
 N.~Pilipenko\thanksref{Peters}
 \and
 A.~Pocar\thanksref{UMass}
 \and
 G.~Raikov\thanksref{Kurchatov}
 \and
 M.T.~Ranalli\thanksref{LNGS}
 \and
 G.~Ranucci\thanksref{Milano}
 \and
 A.~Razeto\thanksref{LNGS}
 \and
 A.~Re\thanksref{Milano}
 \and
  N.~Rossi\thanksref{LNGS}
 \and
 S.~Sch\"onert\thanksref{Munchen}
 \and
 D.~Semenov\thanksref{Peters}
 \and
 G.~Settanta\thanksref{ISPRA,Juelich}
 \and
 M.~Skorokhvatov\thanksref{Kurchatov,Kurchatovb}
 \and
 A.~Singhal\thanksref{GSI,Juelich,RWTH}
 \and
 O.~Smirnov\thanksref{Dubna}
 \and
 A.~Sotnikov\thanksref{Dubna}
 \and
 R.~Tartaglia\thanksref{LNGS}
 \and
 G.~Testera\thanksref{Genova}
 \and
 E.~Unzhakov\thanksref{Peters}
 \and
 A.~Vishneva\thanksref{Dubna}
 \and
 R.B.~Vogelaar\thanksref{Virginia}
 \and
 F.~von~Feilitzsch\thanksref{Munchen}
 \and
 M.~Wojcik\thanksref{Krakow}
 \and
 M.~Wurm\thanksref{Mainz}
 \and
 S.~Zavatarelli\thanksref{Genova}
 \and
 K.~Zuber\thanksref{Dresda}
 \and
 G.~Zuzel\thanksref{Krakow}
 \\ (The Borexino collaboration){}
 }
 
 \thankstext{MPIH}{Present address: Max-Planck-Institut f\"ur Kernphysik, 69117 Heidelberg, Germany}
\thankstext{IHEP}{Present address: IHEP Institute of High Energy Physics, 100049 Beijing, China}
\thankstext{LNGSG}{Present address: INFN Laboratori Nazionali del Gran Sasso, 67010 Assergi (AQ), Italy}
\thankstext{Bicocca}{Present address: Dipartimento di Fisica, Universita degli Studi e INFN Milano-Bicocca, 20126 Milano, Italy}
\thankstext{CALI}{Present address: Department of Physics and Astronomy, University of California, Irvine, California, USA}
\thankstext{GSIMAINZ}{Present address: GSI Helmholtzzentrum f\"ur Schwerionenforschung GmbH, Planckstr. 1, 64291 Darmstadt, Germany and Institute of Physics and EC PRISMA$+$, Johannes Gutenberg Universit\"at Mainz, Mainz, Germany}
\thankstext{MI}{Dipartimento di Fisica, Universit\`a degli Studi e INFN, 20133 Milano, Italy}
\thankstext{BOS}{Present address: Boston University, College of Arts and Sciences, Department of Physics, 02215 Boston, MA, USA}
\thankstext{ISPRA}{Present address: Istituto Superiore per la Protezione e la Ricerca Ambientale, 00144 Roma, Italy}
\thankstext{GSI}{Present address: GSI Helmholtzzentrum f\"ur Schwerionenforschung GmbH, Planckstr. 1, 64291 Darmstadt, Germany}

 \institute{Dipartimento di Fisica, Universit\`a degli Studi e INFN, 20133 Milano, Italy\label{Milano}
 \and
 Chemical Engineering Department, Princeton University, Princeton, NJ 08544, USA\label{PrincetonChemEng}
 \and
  INFN Laboratori Nazionali del Gran Sasso, 67010 Assergi (AQ), Italy\label{LNGS}
 \and
  Physics Department, Princeton University, Princeton, NJ 08544, USA\label{Princeton}
 \and
 Dipartimento di Fisica, Universit\`a degli Studi e INFN, 16146 Genova, Italy\label{Genova}
 \and
  Lomonosov Moscow State University Skobeltsyn Institute of Nuclear Physics, 119234 Moscow, Russia\label{Lomonosov}
 \and
   St. Petersburg Nuclear Physics Institute NRC Kurchatov Institute, 188350 Gatchina, Russia\label{Peters}
 \and
 National Research Centre Kurchatov Institute, 123182 Moscow, Russia\label{Kurchatov}
 \and
  APC, Universit\'e de Paris, CNRS, Astroparticule et Cosmologie, Paris F-75013, France\label{APC}
 \and
  Gran Sasso Science Institute, 67100 L'Aquila, Italy\label{GSSI}
 \and
  Joint Institute for Nuclear Research, 141980 Dubna, Russia\label{Dubna}
 \and
 Institute of Physics and Excellence Cluster PRISMA+, Johannes Gutenberg-Universit\"at Mainz, 55099 Mainz, Germany\label{Mainz}
 \and
   M.~Smoluchowski Institute of Physics, Jagiellonian University, 30348 Krakow, Poland\label{Krakow}
 \and
 Institute for Nuclear Research of NAS Ukraine, 03028 Kyiv, Ukraine\label{Kiev}
 \and
  Department of Physics, Royal Holloway, University of London, Egham, Surrey, TW20 OEX, UK\label{London}
 \and
 Institute of Nuclear Research (Atomki), Debrecen, Hungary\label{Atomki}
 \and
 Institut f\"ur Kernphysik, Forschungszentrum J\"ulich, 52425 J\"ulich, Germany\label{Juelich}
 \and
    RWTH Aachen University, 52062 Aachen, Germany\label{RWTH}
 \and
  National Research Nuclear University MEPhI (Moscow Engineering Physics Institute), 115409 Moscow, Russia\label{Kurchatovb}
 \and
 Physik-Department, Technische Universit\"at  M\"unchen, 85748 Garching, Germany\label{Munchen}
 \and
  Dipartimento di Chimica, Biologia e Biotecnologie, Universit\`a degli Studi e INFN, 06123 Perugia, Italy\label{Perugia}
 \and
 Amherst Center for Fundamental Interactions and Physics Department, UMass, Amherst, MA 01003, USA\label{UMass}
  \and
 Physics Department, Virginia Polytechnic Institute and State University, Blacksburg, VA 24061, USA\label{Virginia}
 \and
 Department of Physics, Technische Universit\"at Dresden, 01062 Dresden, Germany\label{Dresda}
  }
\date{Received: date / Accepted: date}
\maketitle

\begin{abstract}
A search for solar axions and axion-like particles produced in the $p+d\rightarrow\rm{^3He}+A~(5.5\rm{ ~MeV})$ reaction was performed using the complete dataset of the Borexino detector ($3995$ days of measurement live-time). 
The following interaction processes have been considered: axion decay into two photons $({\rm A}\rightarrow2\gamma)$, inverse Primakoff conversion on nuclei $({\rm A}+Z\rightarrow\gamma+Z$), the Compton conversion of axions to photons $({\rm A}+e\rightarrow e+\gamma)$ and the axio-electric effect $({\rm A}+e+Z\rightarrow e+Z$). 
Model-independent limits on product of axion-photon ($g_{A\gamma}$), axion-electron ($g_{Ae}$), and isovector axion-nucleon ($g_{3AN}$) couplings are obtained: $|g_{A\gamma}\times g_{3AN}| \leq 2.3\times 10^{-11} \rm{GeV}^{-1}$ and $|g_{Ae}\times g_{3AN}| \leq 1.9\times 10^{-13}$  at $m_A <$ 1 MeV (90\% c.l.).
The Borexino results exclude new large regions of $g_{A\gamma}$, and $g_{Ae}$ coupling constants and axion masses $m_A$, and leads to constraints on the products $|g_{A\gamma}\times m_A|$ and $|g_{Ae}\times m_A|$ for the KSVZ- and the DFSZ-axion models.
\end{abstract}

\keywords{axion, pseudoscalar particles; low background measurements}

\section{Introduction}\label{intro}
Axions and axion-like particles are widely discussed in modern particle physics~\cite{PDG24}.
The reason for this is that axions not only solve the problem of the apparent absence of CP-violating effects in Quantum Chromodynamics (QCD), but are also very suitable candidates for the role of dark matter particles.
The searches for heavy axions in the keV-MeV-scale mass range are supported by their significance in astrophysics and cosmology.
These particles can affect on the evolution of low-mass stars~\cite{Luc22} and the low-energy Supernovae~\cite{Cap22}, the Big Bang Nucleosynthesis~\cite{Mul23} and the Cosmic Microwave Background~\cite{Cad11,Dep20}.

Originally the axion hypothesis was introduced by Weinberg~\cite{Wei78} and Wilczek~\cite{Wil78}, who showed that the solution of the strong CP-problem proposed earlier by Peccei and Quinn~\cite{Pec77}, should lead to the appearance of a new neutral pseudoscalar particle. 
The original PQWW-axion model produced specific predictions for the coupling constants of axions with photons ($g_{A\gamma}$), electrons ($g_{Ae}$), nucleons ($g_{AN}$) and also for axion mass ($m_A$).
The model was soon excluded by numerous experiments with radioactive sources, nuclear reactors and accelerators~\cite{PDG24}.

Soon other benchmark models of ``invisible'' axion which feebly couples to the Standard Model (SM) particles have been proposed, such as the Kim-Shifman-Vainshtein-Zakharov (KSVZ) model~\cite{Kim79,Shi80} and the Dine-Fischler-Srednicki-Zhitnitsky (DFSZ) model~\cite{Zhi80,Din81}. 

The axion mass in these models is determined by the scale of PQ-symmetry violation or the axion decay constant $f_A$:
\begin{equation}\label{ma}
  m_A\approx (f_\pi m_\pi /f_A) (\sqrt{z}/(1+z)),
\end{equation}
where $m_\pi$ and $f_\pi$ are the mass and decay constant of $\pi^0-$meson and  $z = m_u/m_d$ is $u$ and $d$ quark-mass ratio. 
Taking into account the higher order corrections to the axion mass, the equation (\ref{ma}) can be rewritten as: $m_A(\rm{eV})\approx 5.69(5)\times 10^6/\it{f_A} {\rm{(GeV)}}$ \cite{Gri16,Gor2019}. 
The scale of $f_A$ in models of invisible axion is arbitrary and can be extended to the Planck mass. 
Since the amplitude of axion interaction with photons, electrons and nucleons  are proportional to the axion mass and inversely with $f_A$, the interaction between invisible axions and matter can be strongly suppressed.

Given that generic axion interactions scale approximately with $f_A^{-1}$ the effective coupling constants $g_{A\gamma}$, $g_{Ae}$, and $g_{AN}$ appear to be model dependent. 
For example, the KSVZ axion cannot interact directly with leptons, and the constant $g_{Ae}$ exists only because of radiative corrections. 
Also, the constant $g_{A\gamma}$ can differ by more than two orders of magnitude from the values accepted in the KSVZ and DFSZ models~\cite{Kap85}.

The results from present-day experiments are interpreted within these two most popular axion models. 
The main experimental efforts are focused on searching for an axion with a mass in the range of $10^{-6}$ to $10^{-2}$~eV. 
This range is free of astrophysical and cosmological constraints, and relic axions with such a mass are considered to be the most likely dark matter candidates.

Axions have been intensively searched for using a wide range of experimental methods that exploit $g_{A\gamma}$ coupling: solar helioscopes, dark matter haloscopes, photon regeneration and interferometry experiments~\cite{PDG24,Sik20}. 
Additionally there are a variety of collider and beam dump experiments that can search for axion decays or axion bremsstrahlung~\cite{Cap23}.

Dark matter direct detection experiments have been used to search for axioelectric effect which is due to axion-electron coupling constant for solar and relic axions~\cite{Apr22,Agn23}.

Solar axions produced in nuclear magnetic transitions are searched using the resonant absorption by nuclei, that provides restrictions on axion-nucleon couplings \cite{Der23}. 
The same reaction for solar axions with continuous flat spectrum are sensitive to the product $|g_{AN}\times g_{A\gamma}|$ and $|g_{AN}\times g_{Ae}|$~\cite{Abd20}.

Astrophysical data provide complementary strong constraints on the axions parameter space \cite{Cap24}. 
The results of laboratory searches for the axion as well as astrophysical and cosmological axion bounds can be found in \cite{PDG24}.

The goal of this study is to search for solar axions with an energy of $5.5$~MeV,  produced in the $p + d \rightarrow\rm{^3He}+ A$ ($5.49$~MeV) reaction using the large neutrino scintillation the Borexino detector. 
The axion flux is thus proportional to the $pp$-neutrino flux, which is known with a high accuracy~\cite{Ser09,Vin17,Bel11A,Ago19}. 

The range of axion masses under study is expanded more than $5$~MeV, which is interesting because new solutions to the $CP$ problem rely on the hypothesis of a world of mirror particles~\cite{Bere00,Bere01} and super-symmetry~\cite{Hal04} and allow the existence of axions with a mass of about $1$~MeV and reduced $g_A\gamma$ coupling constant.
These axions are not excluded by laboratory experiments or astrophysical data.

The axion detection signatures exploited in this study are axion decay into two $\gamma$-quanta and inverse Primakoff conversion on nuclei, ${\rm A}+Z\rightarrow\gamma+Z$. 
The amplitudes of these reactions depend on the axion-photon coupling $g_{A\gamma}$.
We also consider the potential signals from  Compton axion to photon conversion, ${\rm A}+e\rightarrow e+\gamma$, and from the axio-electric effect, ${\rm A}+e+Z\rightarrow e+Z$. 
The amplitudes of these processes are defined by the $g_{Ae}$ coupling. 
The signature of all these reactions is $5.5$~MeV peak.

We have previously published the result of searches for solar axions emitted in the $478$~keV M1-transition of $^7{\rm{Li}}$ using the Borexino counting test facility~\cite{Bel08} and $5.5$~MeV solar axions using 536$\times$100 days$\times$tons of Borexino detector data~\cite{Bel12a}.

\section{Flux of 5.5 MeV axions from the Sun}
Stars, like our Sun, should be intense sources of axions. 
Intense fluxes of axions can be formed in the Sun in a number of processes whose probabilities depend on the axion coupling constants $g_A\gamma$, $g_{Ae}$, and $g_{AN}$. 
The constant $g_{A\gamma}$ specifies the probability of conversion of photons to axions in the electromagnetic field of the solar plasma (Primakoff axions).
The constant $g_{Ae}$ specifies the axion fluxes from bremsstrahlung and Compton process, as well as in discharge and recombination processes in atoms.
The constant $g_{AN}$ determines the emission of axions in nuclear magnetic transitions from levels that are thermally excited at high temperatures in the center of the Sun.

The nuclear reactions of the $\emph{pp}$-solar chain and the CNO cycle also can produce axions. 
The most intense flux is expected from the reaction of the $\rm{^3He}$ production:
\begin{equation}
 p + d\rightarrow \rm{{^3He} + \gamma ~(5.5\;MeV)}.
\end{equation}

The value of $5.5$~MeV solar axion flux can be expressed in terms of the $pp$-neutrino flux because $99.7$\% of deuterium is produced from the fusion of two protons reaction: $p + p \rightarrow d + e^+ + \nu_e$ . 
The proportionality factor between the axion and $pp$-neutrino fluxes is determined by a ratio between  the probability of nuclear transition with axion production $(\omega_A)$ and photon production $(\omega_\gamma)$.
The ratio $(\omega_A/\omega\gamma)$  depends on dimensionless axion-nucleon coupling constant $g_{AN}$, which consists of isoscalar $g_{0AN}$ and isovector $g_{3AN}$ components.
The isoscalar and isovector parts of the axion–nucleon coupling constant, respectively, are model dependent. 
They can be expressed in terms of the effective axion–proton and axion–neutron coupling constants $C_p$ and $C_n$, respectively~\cite{Cor16,Avi18}:
\begin{eqnarray}
\label{gan0}
g_{0AN}=(M_N/2f_A)(C_p+C_n),\\
g_{3AN}= (M_N/2f_A)(C_p-C_n),
\label{gan3}
\end{eqnarray}
where $m_N\approx939$~MeV is the mass of the nucleon. 
The effective coupling constants $C_p$ and $C_n$ in turn depend on the axion–quark coupling constants \cite{PDG24,Cor16}. 
The calculations performed in \cite{Cor16} give the values $C^{KSVZ}_p=-0.47(3)$ and $C^{KSVZ}_n=0.02(3)$ for KSVZ-model.
As mentioned above, the axion–neutron coupling in KSVZ model is strongly suppressed. 

In the DFSZ-model, $C_p$ and $C_n$ depend on an additional parameter $\beta^*$ which is the ratio of the vacuum expectation values of the two Higgs doublets giving masses to the up- and down-quarks: $\rm{tan}\beta^* = \upsilon_u/\upsilon_d$.
The values $C^{DFSZ}_p=-(0.435~\rm{sin^2}\beta^*+0.182)\pm0.025$ and $C^{DFSZ}_n=(0.414~\rm{sin^2}\beta^*-0.160)\pm0.025$ for the DFSZ axion were obtained in \cite{Cor16} in terms of the corresponding model-dependent quark couplings $C_u=C_c=C_t=1/3~\rm{cos}^2\beta^*$ and $C_d=C_s=C_b=1/3~\rm{sin}^2\beta^*$.
The possible values of $\rm{tan}\beta^*$ are restricted by the range $0.25 \leq \rm{tan}\beta^* \leq 140$ \cite{Che13,Luz20}.
Using Eqs.~(\ref{ma}, \ref{gan0}, \ref{gan3}), one can express isoscalar $g_{0AN}$ and isovector $g_{3AN}$ constants in terms of the axion mass $m_A$.

\begin{figure}
\includegraphics[width=9cm,height=10.5cm]{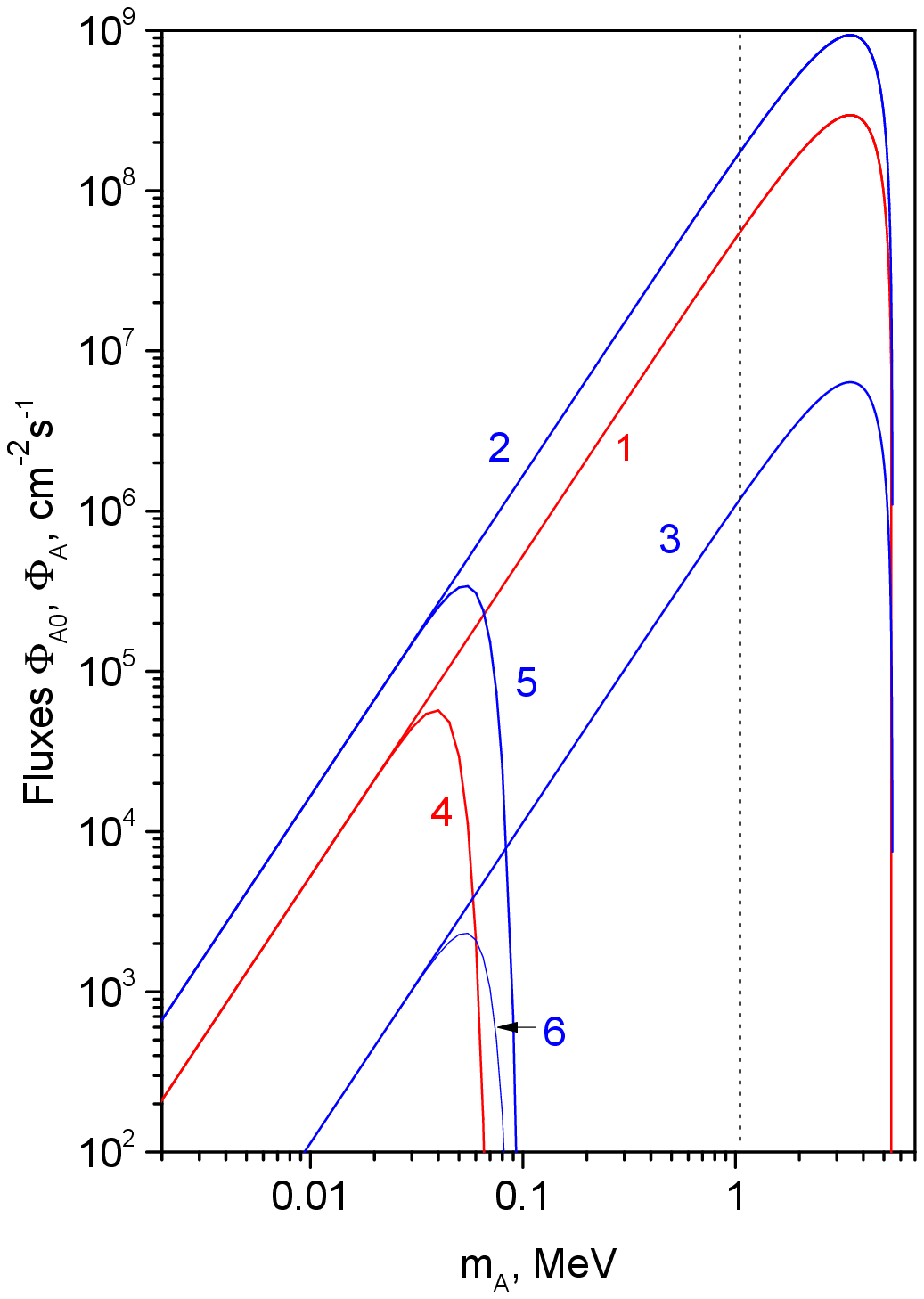}
\caption {The expected fluxes of the solar axions emitted in the $p + d \rightarrow {^3{\rm{He}}} + A$ reaction. 1 - KSVZ, 2 - DFSZ ($\rm{tan}\beta^*=140$), 3 - DFSZ ($\rm{tan}\beta^*=0.25$), 4, 5, 6  - the decay $A\rightarrow2\gamma$ is taken into account, in the KSVZ- and DFSZ-models, respectively. At $m_A \geq 2m_e$ the decay $A\rightarrow e^+e^-$ is possible (dot line).}\label{Figure:Fluxes}
\end{figure}

The M1-type transition in the $p+d\rightarrow \rm{^3He}+\gamma$ reaction corresponds to the capture of a proton from the $S$ state. 
The probability, $\chi$, of proton capture with zero orbital moment at energies below 80 keV was measured in \cite{Sch97}; at a proton energy of $\sim 1$ keV, $\chi$ = 0.55. 
The proton capture from the $S$ state corresponds an isovector transition, and the ratio $\omega_A/\omega_\gamma$ depends only on $g_{3AN}$ \cite{Don78,Hax91,Avi88}:

\begin{equation}\label{ratio}
\frac{\omega_{A}}{\omega_{\gamma}} =
 \frac{\chi}{2\pi\alpha}\left[\frac{g_{3AN}}{\mu_3}\right]^2\left(\frac{p_A}{p_\gamma}\right)^3 = 0.54(g_{3AN})^2 \left(\frac{p_A}{p_\gamma}\right)^3,
\end{equation}

where $p_{\gamma}$ and $p_{A}$ are, respectively, the photon and axion momenta, $\alpha\approx 1/137$ is the fine-structure constant, $\mu_3= \mu_p-\mu_n\approx4.71\mu_N$ is the isovector nuclear magnetic moment in
units of nuclear magneton and $g^{KSVZ}_{3AN} = -0.49~(M_N/2f_A)$ and $g^{DFSZ}_{3AN} = -(M_N/2f_A)(0.849~\rm{sin^2}\beta^*+0.022)$ for KSVZ- and DSVZ-models, respectively.

The expected solar axion flux on the Earth's surface is then
\begin{eqnarray}\label{FluxA}
\Phi_{A0} = \Phi_{\nu pp}(\omega_A/\omega_\gamma) = 3.23\times10^{10}(g_{3AN})^2(p_A/p_\gamma)^3,
\end{eqnarray}
where $\Phi_{\nu p p} = 6.0 \times 10^{10} {\rm{cm}}^{-2} {\rm{s}}^{-1}$ is the $pp$ solar neutrino flux \cite{Ser09,Vin17,Bel11A,Ago19}.
Using the relation between $g_{3AN}$ and $m_A$ given by (\ref{gan3}), the $\Phi_{A0}$ value appears to be proportional to $m_A^2$: $\Phi_{A0} = 5.28\times10^{-5}m_A^2(p_A/p_\gamma)^3~\rm{cm^{-2}~s^{-1}}$ (KSVZ), $\Phi_{A0} = 16.7\times10^{-5}m_A^2(p_A/p_\gamma)^3~\rm{cm^{-2}~s^{-1}}$ (DFSZ, $\rm{tan}\beta^*=140$), and $\Phi_{A0} = 1.14\times10^{-6}m_A^2(p_A/p_\gamma)^3~\rm{cm^{-2}~s^{-1}}$ (DFSZ, $\rm{tan}\beta^*=0.25$), where $m_A$ is given in eV units.
Apparently, the DFSZ-axion flux can be more than two orders of magnitude lower for certain values of the $\beta^*$-parameter (see Fig.~\ref{Figure:Fluxes}).

The dependence of the flux $\Phi_{A0}$ on the axion mass $m_A$ has a bell-shaped form reaching a maximum at an energy of $3.5$~MeV, for small $m_A$ the flux is suppressed by the factor $m_A^2$ and for large $m_A$ by the factor $(p_A/p_\gamma)^3$.
The calculated values of the uncharged  fluxes $\Phi_{A0}(m_A)$ as a function of the axion mass are shown in Fig.~\ref{Figure:Fluxes} for KSVZ- and DSVZ-models (lines 1, 2, and 3).
\section{Main reactions of 5.5~MeV axion in the Borexino} 
\subsection{Axion decay $\rm{A}\rightarrow 2\gamma$ and axion conversion on nuclei ${\rm A}+\rm{Z}\rightarrow\gamma+\rm{Z}$}

As was shown above, the flux of axions with an energy of $5.5$~MeV is determined by the isovector coupling constant of the axion with nucleons $g_{3AN}$.
Four different reactions were analysed for the detection of $5.5$~MeV axions.
The decay of an axion into two photons and the conversion of an axion into a photon on nuclei are determined by the coupling constant of the axion to the photon $g_{A\gamma}$.
The axion-photon coupling constant $g_{A\gamma}$ is presented in~\cite{Gri16,Kap85,Sre85}:
\begin{equation}
g_{A\gamma}=\frac{\alpha}{2\pi f_{A}}\left(\frac{E}{N}-\frac{2(4+z)}{3(1+z)}\right)\equiv\frac{\alpha}{2\pi
f_{A}}C_{A\gamma\gamma}\label{C_gamma}
\end{equation}
where $E/N$ is a model dependent parameter of the order of unity.
$E/N = 8/3$ in the DFSZ axion models ($C_{A\gamma\gamma}$=0.74) and $E/N = 0$ for the original KSVZ axion ($C_{A\gamma\gamma} = -1.92$).
There are QCD axion models whose $g_{A\gamma}$ couplings constant are outside the region restricted  by this two values, motivating searches for axions over a wide range of masses and couplings constants.

The decay $\rm{A}\rightarrow 2\gamma$ was the first reaction used for experimental search for standard PQWW-axion.
The mean lifetime of the decay is given by the expression:
\begin{equation}
\tau_{2\gamma}=\frac{64\pi \hbar}{g_{A\gamma}^2m_A^3}.\label{tau_CM}
\end{equation}
In numerical form, for $\tau_{2\gamma}$ measured in seconds, $g_{A\gamma}$ in ${\rm{GeV^{-1}}}$, and $m_A$ in ${\rm{eV}}$, one obtains:
\begin{equation}
\tau_{2\gamma}~(s)=1.3\times 10^5 g_{A\gamma}^{-2}m_A^{-3} = 3.5\times 10^{24}m_A^{-5}C_{A\gamma\gamma}^{-2}. \label{axion_decay}
\end{equation}
The flux of solar axions reaching the Earth is given by
\begin{equation}
\Phi_{A}=\Phi_{A0}{\rm{exp}}(-\tau_{f}/\tau_{2\gamma}) = \Phi_{A0}{\rm{exp}}(-\tau_{f} g_{A\gamma}^2 m_A^3/64\pi) \label{Flux}
\end{equation}
where $\Phi_{A0}$ is the axion flux at the Earth in case there is no axion decay (\ref{FluxA}), $\tau_{2\gamma}$ is defined by (\ref{tau_CM},
\ref{axion_decay}), and $\tau_{f}$ is the time of flight in the axion frame of reference.
Because of axion decay, the sensitivity of experiments searching solar axions drops off for large values of $g_{A\gamma}^{2}m_A^{3}$.
The expected axion spectra for KSVZ- and DFSZ-models as function of $m_A$ are shown in~Fig.~\ref{Figure:Fluxes} (lines 4 KSVZ-, and lines 5 and 6 for DFSZ-axions). 
The fluxes have a bell-shaped form with maximum at $\sim(40-50)$~keV since at low $m_A$ the probability of decay is low and at large ones the axion flux is reduced due to decays $\rm{A}\rightarrow 2\gamma$ on the way to the Earth.

The results were interpreted in terms of QCD-axion models using relevant relations between the axion mass and the coupling constants.
In case of axion-like particles, the QCD-axion relations no longer apply and the results must be analyzed with axion mass and couplings as independent parameters.

The inverse Primakoff conversion ${\rm A}+Z\rightarrow\gamma+Z$ is similar to well known $\pi^0-$meson to photon conversion on nuclei.
In the case of the Borexino, the conversion on the carbon nuclei is the most probable one:
$A+{^{12,13}\rm{C}}\rightarrow\gamma+{^{12,13}\rm{C}}$. 
The integral cross section of inverse Primakoff conversion is given in~\cite{Avi88}:
\begin{equation}
\sigma_{PC}=g_{A\gamma}^{2}\frac{Z^{2}\alpha}{2}\left[\frac{1+\beta^{2}}{2\beta^{2}}\ln\left(\frac{1+\beta}{1-\beta}\right)-\frac{1}{\beta}\right].\label{Primakoff}
\end{equation}
The cross section (\ref{Primakoff}) depends on the $g_{A\gamma}$ coupling and the decrease of the axion flux due to $A\rightarrow 2\gamma$ decays during their flight from the Sun should be taken into account. 
The maximum count rate, as for the case of $A\rightarrow 2\gamma$ decay is expected for axion masses of about 50~keV.

\begin{figure}
\includegraphics[width=8.5cm,height=10.5cm]{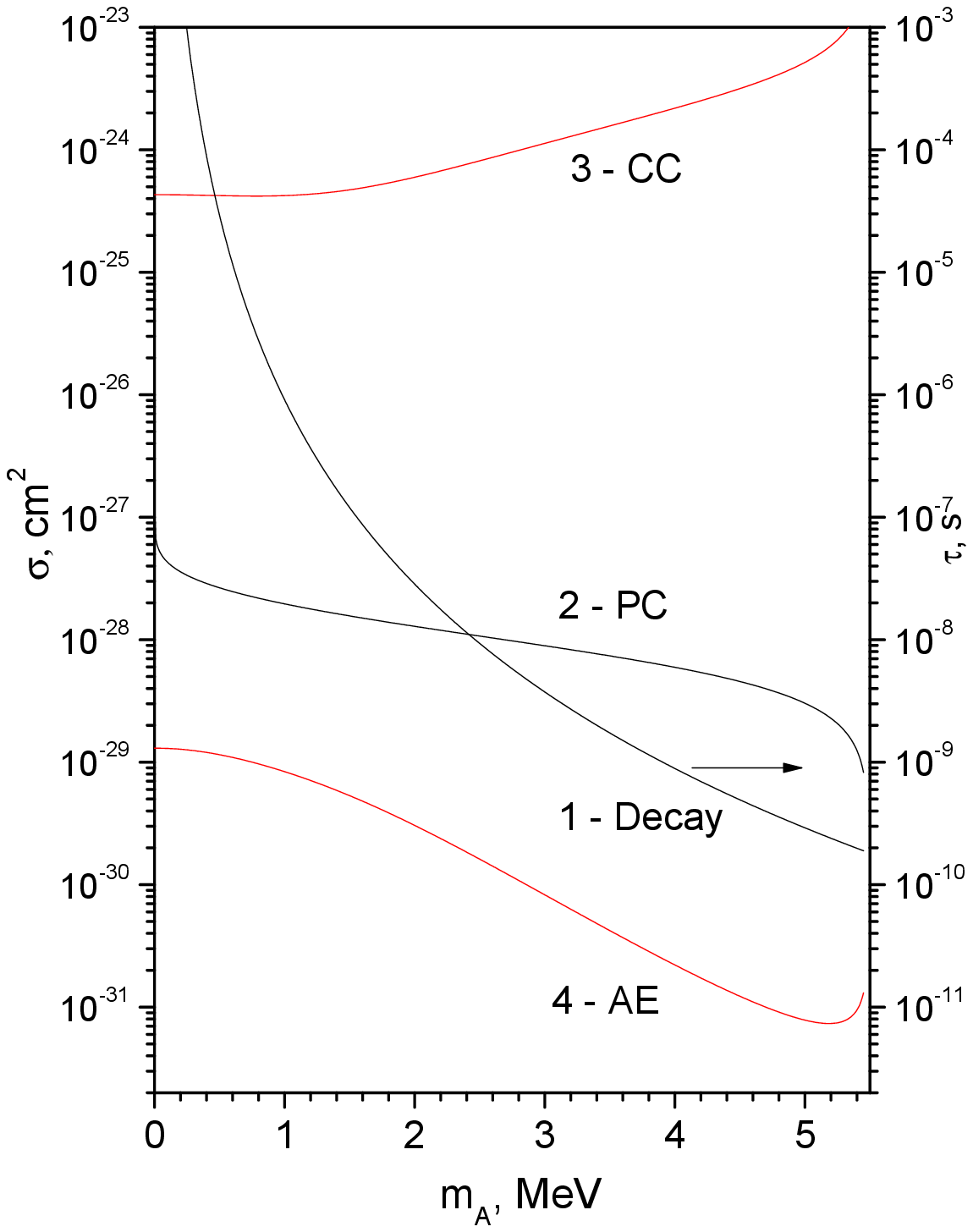}
\caption {The cross-section of axion reactions and axion lifetime vs axion mass. 1 - axion lifetime for $A\rightarrow 2\gamma$ decay (the vertical scale shows $\tau\times10^{-15}$~s), 2 - inverse Primakoff conversions on $^{12}\rm{C}$ nuclei, 3 -  Compton conversion and 4 - axioelectric effect on carbon atoms. The lifetime and cross-section of 1 and 2 were calculated for $g_{A\gamma}=1~\rm{GeV^{-1}}$, the cross-sections of 3 and 4 with $g_{Ae}=1$.}
\label{Figure:CrossSections}
\end{figure}

\subsection{Compton conversion $A+e\rightarrow\gamma+e$ and axio-electric effect $A+e+Z\rightarrow e+Z$}
Two other studied reactions - the Compton process ${\rm A}+e\rightarrow e+\gamma$, and the axioelectric effect ${\rm A}+e+Z\rightarrow e+Z$ - depend on the coupling constant between the axion and the electron $g_{Ae}$.

The dimensionless coupling constant $g_{Ae}$ is associated with the electron mass $m_e$, so that $g_{Ae}=C_em_e/f_{A}$, where $C_e$ is a model dependent factor of the order of unity. 
In the standard PQWW axion model, the values $f_A$=250 GeV and $C_e$=1 are fixed and $g_{Ae}\approx 2\times10^{-6}$. 
In the DFSZ axion models, $C_e=\sin^2\beta^*/3$, where as was mentioned above $\beta^*$ is an unknown angle, such that $\tan\beta^*=\upsilon_u/\upsilon_d$ is the ratio of the vacuum values of the two Higgs doublets. 
Assuming e.g. $\tan\beta^*$=10 \cite{Avi18}, the axion-electron coupling is $g_{Ae}$=2.96$\times10^{-11}m_A$ where $m_A$ is expressed in $eV$ units. 

The hadronic axion has no tree-level couplings to the electron, but there
is an induced axion-electron coupling at one-loop level \cite{Sre85}:
\begin{equation}
g_{Ae}=\frac{3\alpha^{2}m_e}{4\pi^2 f_{A}}\left(\frac{E}{N}\ln\frac{f_{A}}{m_e}-\frac{2}{3}\frac{(4+z)}{(1+z)}\ln\frac{\Lambda}{m_e}\right)\label{Gaee} 
\end{equation}
where 
$\Lambda\approx$1 GeV is the cutoff at the QCD confinement scale and the ratio $2(4+z)/3(1+z)\approx1.92$. 
In the KSVZ model, $E/N =0$, the first term drops out and the expression (\ref{Gaee}) can be represented as depending on the axion mass $g_{Ae}=5.29\times10^{-15}m_A/1 \rm{eV}$.
The interaction strength of the hadronic axion with the electron is suppressed by a factor $\sim\pi\alpha^2$.

In the Compton-like process $A+e\rightarrow\gamma+e$ axion is scattered by electron with $\gamma$-quanta production.
The spectrum of the photons depends on the axion mass, while the spectra of electrons can be found from relation $E_e = E_A-E_{\gamma}$, where $E_A\cong$ 5.49 MeV. 
The cross section of the reaction is proportional $g_{Ae}^2$ and was calculated in  \cite{Don78,Avi88,Zhi79,Bel12a}.
The cross section has a complex form, the phase space contribution is approximately independent of $m_{A}$ for $m_{A}<$ 2 MeV and the integral cross section is:
\begin{equation}
\sigma_{CC}\approx g_{Ae}^{2}\times4.3\times10^{-25} \rm{cm^{2}}.\label{CC_CS}
\end{equation}

The axioelectric effect $A+e+Z\rightarrow e+Z$ is the analogue of the photoelectric effect and depends on $g_{Ae}$-coupling. 
The cross-sections of the axioelectric effect, expressed through the cross-section of the photoelectric effect, including for the case of non-relativistic axions, were obtained in work \cite{Pos08}.
The cross section of the axio-electric effect on K-electrons where the axion energy $E_{A}\gg E_{b}$ was calculated in \cite{Zhi79} and has a complex form also.  
The cross section has a $Z^{5}$ dependence and for carbon atoms the cross section is $\sigma_{Ae}\approx g_{Ae}^{2}\times$1.3$\times$10$^{-29}$ cm$^{2}$/electron for $m_A <$ 1 MeV.

This value is 4 orders of magnitude lower than for axion Compton conversion. 
However, thanks to the different energy dependence ($\sigma_{CC}\sim E_{A}$,
$\sigma_{Ae}\sim(E_{A})^{-3/2}$) and $Z^{5}$ dependence, the axio-electric effect is a potential signature for axions with detectors having high $Z$ active mass \cite{Der10,Der13,Der14}.

The cross-sections of Primakoff conversion $\sigma_{PC}$ on isotope $\rm{^{12}C}$, Compton conversion  $\sigma_{CC}$ and axio-electic effect  $\sigma_{AE}$ on carbon atoms calculated fo $g_{A\gamma}=1~\rm{GeV^{-1}}$ and $g_{Ae}=1$  are shown in Fig.\ref{Figure:CrossSections}. 
Here is also shown the lifetime of $A\rightarrow2\gamma$ decay in seconds multiplied by $10^{-15}$ to match the values on the left scale.

The requirement that most axions escape the Sun limits above the possible axion coupling strengths.
The analysis performed in our previous paper~\cite{Bel12a} showed that the sensitivity of 5.5 MeV solar experiments is limited by the following values of coupling constants: $g_{A\gamma} < 10^{-4}~\rm{GeV}^{-1}$, $g_{Ae}< 10^{-6}$,  and $g_{AN} < 10^{-3}$.

\section{Borexino response functions and data selection} 

\subsection{Borexino detector}
The design and main components and features of the Borexino is described in detail in \cite{Ago19,Ali02,Ali09,Bac12,Bel14}.
Borexino is a scintillator detector with an active mass of 278 tons of pseudocumene, doped with $1.5$~g/L of PPO (2.5-diphenyloxazole, a fluorescent dye). 
The scintillator is housed in a thin nylon inner vessel (IV) and is surrounded by two concentric pseudocumene buffers ($323$ and $567$~tons) doped with a small amount of light quencher to reduce their scintillation.
The two buffers are separated by a second thin nylon membrane to prevent diffusion of external radon. 
The scintillator and buffers are contained in a stainless steel sphere (SSS) with diameter $13.7$~m. 
The SSS is enclosed in domed water tank (WT), containing 2100 tons of ultra pure water as a shield against external $\gamma$'s and neutrons. 
The scintillation light is detected by $2212$~$8''$~PMTs (PMT - photomultiplier) distributed on the SSS. 
The WT is equipped with $208$ additional PMTs that act as a Cerenkov muon detector (outer detector) to identify cosmogenic muons.
The  energy and spatial resolution of the detector were studied with radioactive sources placed at different points inside the~IV~\cite{Bel10,Bel10A,Bel11d}.

\subsection{Response function to axion reactions}

The signature of  these four reactions discussed above is a $5.5$~MeV peak.
The Monte Carlo (MC) method based on the GEANT4 framework~\cite{All16} was used to simulate the Borexino response to electrons and $\gamma$-quanta produced by axion interactions~\cite{Ago18}. 
The simulations have taken into account the effect of ionization quenching, Cerenkov radiation and small non-linearity induced by the energy dependence on the event position. 

For each of the four reactions $3.5\times10^5$ MC events were simulated uniformly within a sphere with radius of $5$~m that fully covered the IV and uniformly in time in order to follow temporal evolution of the detector condition.
The MC candidate events were selected by the same procedure that was applied in the real data selection. The fraction of axion events ($\epsilon$) surviving this procedure could be found in table~\ref{tab:Efficiency}.

\begin{table}[!htbp] 
\begin{center}
\caption{The fraction $\epsilon$ of axion events surviving data selection procedure. CC - Compton axion to photon conversion, $A+e\rightarrow e+\gamma$; AE - axioelectric effect, $A+e+Z\rightarrow e+Z$; PC - Primakoff conversion on nuclei, $A+{^{12}\rm{C}}\rightarrow\gamma+{^{12}\rm{C}}$.} \label{tab:Efficiency}
\begin{tabular}{|c|c|c|c|c|}
      \hline
     reaction& CC & AE & $A$$\rightarrow$2$\gamma$ & PC    \\ 
     \hline
     $\epsilon$ &  0.251  & 0.225  &  0.247   & 0.175        \\
    \hline
\end{tabular}
\end{center}
\end{table}

The responses for the axion decay into two $\gamma$ quanta were calculated taking into account the angular correlation between photons. 
The energy spectra of electrons and $\gamma$-quanta from the axion Compton conversion were generated according to the differential cross-section given in~\cite{Don78,Avi88,Zhi79} for different axion masses~\cite{Bel08,Bel12a}.

\begin{figure}
\centering
\includegraphics[width=\linewidth]{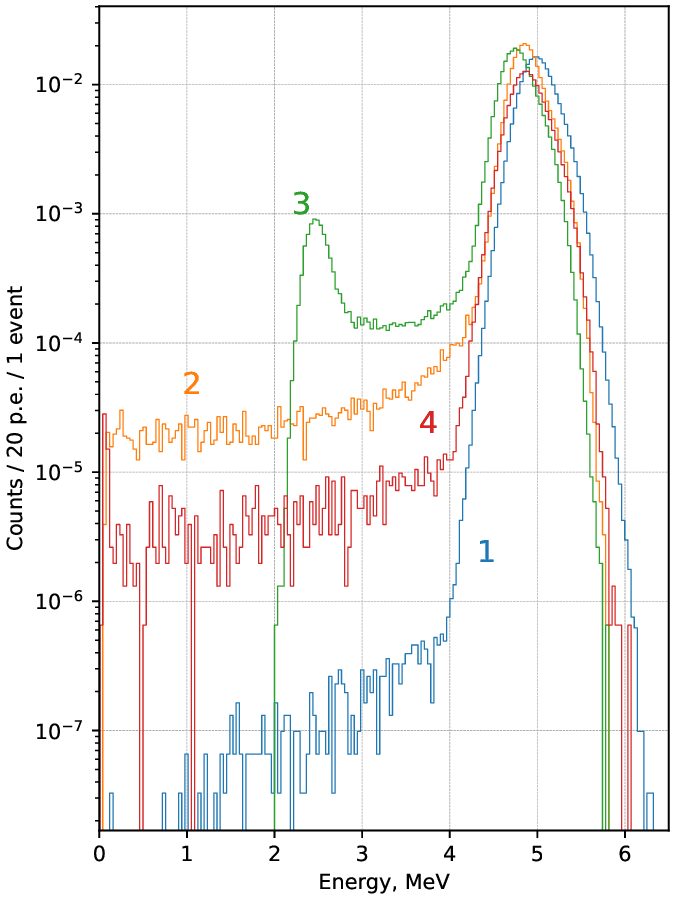}
\caption{\label{Figure:BorexinoResponse}Simulated responses to axion interactions in the Borexino IV: 1 - axioelectric effect ($5.49$~MeV electrons), 2 - Compton axion to photon conversion (electrons and $\gamma$-quanta), 3 - decay $A\rightarrow 2\gamma$,  4 - Primakoff conversion ($5.49$~MeV $\gamma$-quanta). The response functions are normalized to single axion interaction or decay in $5$~m radius sphere.} \label{Figure:Response_Functions}
\end{figure}

The response functions for the axio-electric effect (electron with energy 5.5 MeV), for axion Compton conversion (electron and $\gamma$-quanta with total energy of 5.5 MeV),  axion decay (two $\gamma$-quanta with energy 2.75 MeV in case of non-relativistic axions) and for Primakoff conversion (5.5 MeV $\gamma$-quanta) are shown in Fig.\ref{Figure:Response_Functions}. 

The simulation of response functions as well as the fitting of the measured Borexino spectrum were carried out using the variable Np.e. which is the total number of collected photoelectrons (p.e.) normalized to  the run-dependent number of working PMTs.
The energy of an event is expressed approximately as $E \rm{(MeV)} = Np.e./500$.
The response functions are normalized to one axion interaction (decay) in the sphere with radius of 5~m where the initial events were simulated.

\subsection{Data selection}
The Borexino detector carried out measurements from May 15, 2007 to October 7, 2021.
The aim of data selection is to provide maximum exposure for the desired study with minimum background contribution. 
Background composition of the Borexino experiment was carefully studied in the course of many years of research.  
In the current analysis we aim to the energy region around of 5.5 MeV. 
The background components that have to be taken into account are the short-lived cosmogenic beta-sources ($^{12}\rm{B}$, $^{8}$He, $^{9}$C, $^{9}$Li, etc.), decays of internal $^{208}$Tl, external high-energy gamma produced due to neutron radiative capture in the SSS and solar neutrinos produced in the $^8$B decay. 

The cosmogenic component suppression is performed with a system of temporal and spatial veto that includes:\\
  - total detector veto for $120$~s after each muon shower identified by observation of over $20$ neutron captures in the neutron gate ($1.6$~ms right after the muon event),\\
  - total detector veto for $4$~s after each muon crossing the SSS,\\
  - cylindrical veto with radius of $1$~m for $20$~s on each muon track that crosses the SSS,\\
  - spherical veto with radius of $1$~m and duration of $20$~s for every neutron detected in the muon gate.  

This system of cosmogenic veto allows for maximal suppression of the cosmogenic background for the price of $22.8$\% exposure loss.

Whilst the internal $^{208}$Tl and the solar neutrinos are uniform in the bulk of the detector and can not be suppressed or discriminated, the external high-energy gammas can be discriminated. 
Since these events count rate decreases exponentially towards the center of the detector, they can be suppressed by setting up a fiducial volume. 
In this analysis, the fiducial volume was defined by minimal distance to the inner vessel of $0.75$~m and had the mass of around $145$~tons. 
Moreover, these events were simulated with a full MC simulation following the detector evolution and this simulation was used for a so-called multivariate fit that takes into account the radial distribution providing a discrimination between external and internal spectrum components. 

\begin{figure}
\includegraphics[width=\linewidth]{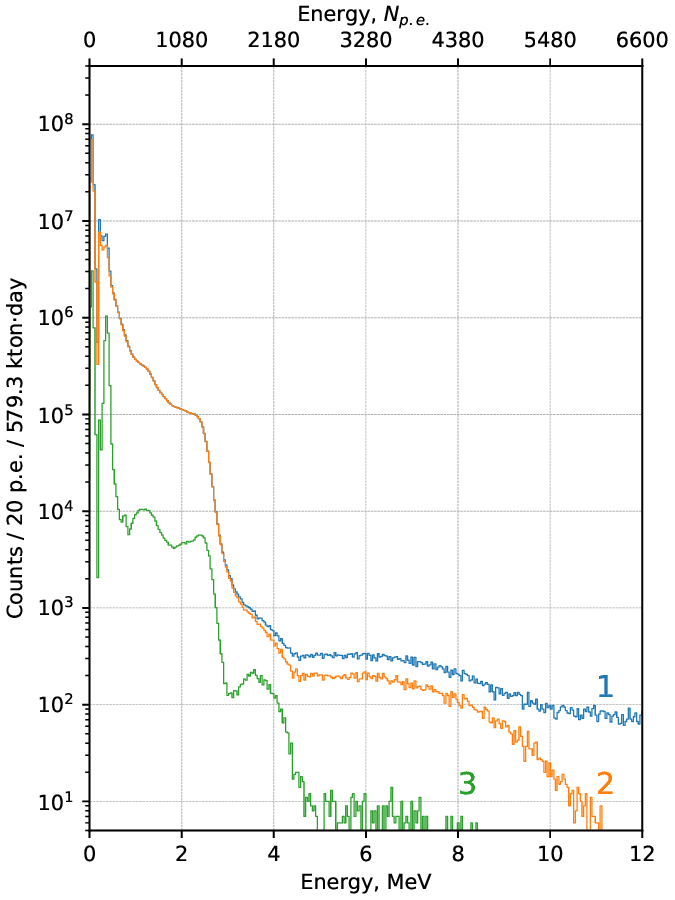}
\caption { Energy spectra of the events and effect of the selection cuts. From top to bottom: 1 - raw spectrum; 2 - with $2$~ms muon veto cut; 3 - final spectrum after application of the cosmogenic veto and FV cut. The bumps at $0.4$, $1.3$, $2.4$, and $3.3$~MeV are caused by decays of $^{210}$Po, $^{11}$C, external $\gamma$'s from $^{208}$Tl and internal $^{208}$Tl, respectively.} \label{Figure:Spectra}
\end{figure}

The experimental energy spectrum from Borexino in the range up to 6600 p.e. ($\simeq12$~MeV), containing $579.3$~kton~days of data, is shown in Fig.~\ref{Figure:Spectra}.  
At energies below $3$~MeV, the spectrum is dominated by $2.6$~MeV $\gamma$'s from the $\beta$-decay of $^{208}$Tl in the PMTs and in the SSS.
The bumps in the spectrum observed at $200$~p.e., $700$~p.e., $1300$~p.e., and $1800$~p.e. are caused by $\alpha$-decays of $^{210}$Po, $\beta^+$-decays of $\rm{^{11}C}$, external $\gamma's$ from $\rm{^{208}Tl}$ and internal $\beta^-$-decays of $\rm{^{208}Tl}$,  respectively (more details e.g. in~\cite{Ago19}).

\section{New limits on axion coupling constant and axion mass}
\subsection{Analysis of the spectrum}
\begin{figure}
\includegraphics[width=\linewidth]{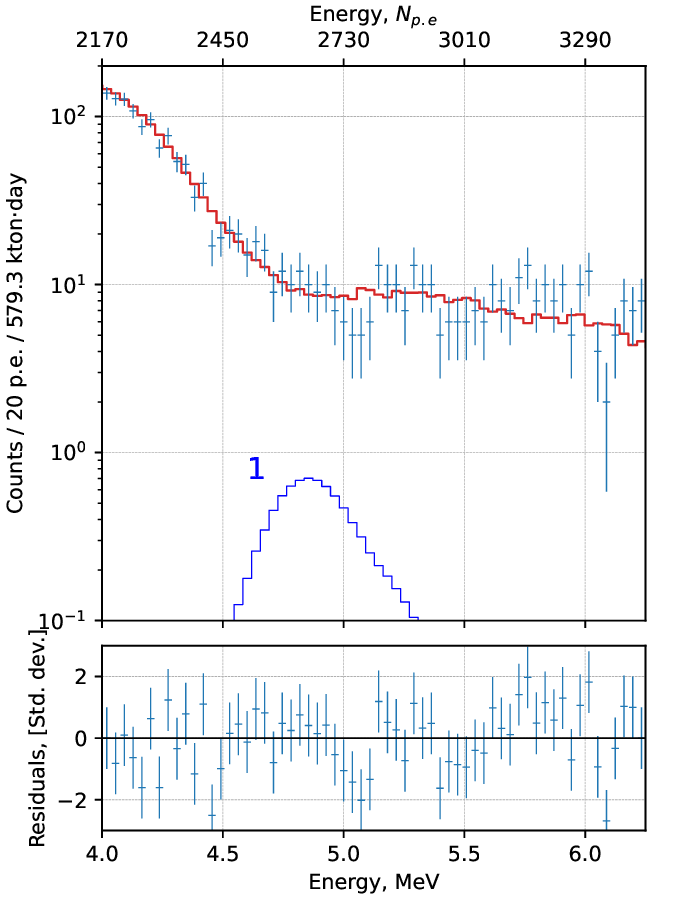}
\caption {Fit of the Borexino energy spectrum in the ($4.4-6.8$)~MeV range. Curve 1 is the detector response function for axio-electric effect on carbon atoms and corresponds to $90$\%~c.l. upper limit ($S = 5.6$~events). The fit (curve 2) is complementary to the radial fit on Fig.~\ref{Figure:Fit_R3}.} \label{Figure:Fit_AE}
\end{figure}

Figure \ref{Figure:Fit_AE} shows the observed Borexino energy spectrum in the ($4.4-6.8$)~MeV ($(2200-3400)$ p.e.) range in which the axion peaks might appear.
The spectrum $N(E)$ was fitted by the sum of four main components: the $\rm{^{208}Tl}$ decays inside the scintillator $N_{int}$, the simulated spectrum of external backgrounds caused by consequent $(\alpha,n)$ reactions followed by neutron captures $(n, \gamma)$ on Ni, Fe, Cr nuclei contained in the SSS constructions $N_{ext}$, the solar $\rm{^8B}$-neutrino-electron scattering events $N_{8B\nu}$, and the simulated spectrum of axion response $N_A$.
\begin{equation}
   N(E) = N_{int}(E)+N_{ext}(E) +N_{8B\nu}(E) + N_A(E),
   \label{eq:Bkg}
\end{equation}
The fitting results are shown in Fig.~\ref{Figure:Fit_AE}. 
The right side of internal $\rm{^{208}Tl}$ decays bump $N_{int}$ was described with the Gaussian function with three free parameters. Such description of this bump was tested on internal $\rm{^{208}Tl}$ MC simulated data and has shown good statistical agreement with it.
The spectrum of recoil electrons due to solar $\rm{^8B}$-neutrino scattering in this interval is well fitted by an exponential function with two free parameters.
As was mentioned, the external background and axion spectra were simulated, only the number of events in the spectrum were free in the fit.

All background components in~(\ref{eq:Bkg}) normalised per unit volume have their own characteristic radial dependencies.
The radial dependence of initially uniform  $N_{int}$, $N_{8B\nu}$ and $N_A$ events is mainly due to the spatial cut of events near the detector poles, as well as due to the slight deviations of inner vessel from spherical shape.
The radial distribution of external backgrounds $N_{ext}$ was taken from the full Monte-Carlo simulation.

The energy spectrum and the radial distribution were fitted simultaneously through minimization of the  two histograms $\chi^2$ functions sum. 
The best-fit amplitude of the axion peak was statistically indistinguishable from zero. 
Since the average value of events per bin in the detector data was reasonably large it became possible to derive the limits on the axion interaction rate from the likelihood function profile. 
The obtained upper limits on the number of axion events are shown in the table \ref{tab:Upper_Limits}.

\begin{table}[!htbp] 
\begin{center}
\caption{The upper limits on the number of axions registered in Borexino FV (counts/579 kton-days). CC - Compton axion to photon conversion,
$A+e\rightarrow e+\gamma$; AE - axio-electric effect, $A+e+Z\rightarrow e+Z$; PC - Primakoff conversion on nuclei, $A+{^{12}C}\rightarrow\gamma+{^{12}C}$.
The limits are given at 90$\%$ c.l.}\label{tab:Upper_Limits}
\begin{tabular}{|l|c|c|c|c|}
      \hline
    reaction& CC & AE & $A$$\rightarrow$2$\gamma$ & PC     \\ \hline
     $S^{\rm lim}$    & 8.7  &  5.6   & 15.9   &  10.2       \\
\hline
\end{tabular}
\end{center}
\end{table}

\begin{figure}
    \centering
    \includegraphics[width=\linewidth]{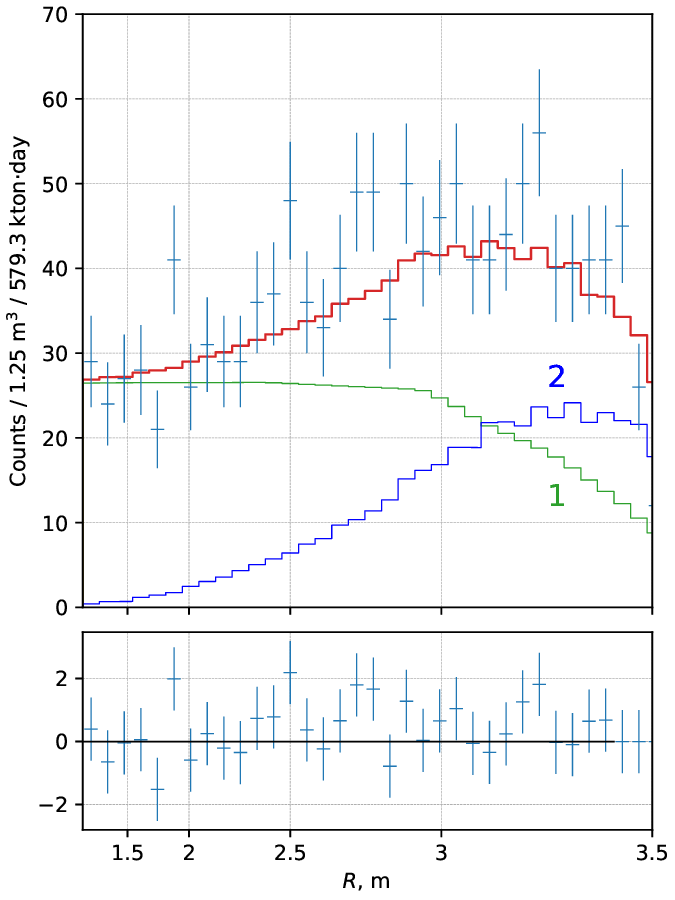}
    \caption{The fitted Borexino radial spectrum. 
    The curve 1 corresponds to the uniform part of the spectrum whilst the curve 2 is the radial distribution of the simulated external background events. The decline in the distribution of uniform events is caused by the fiducial volume non-sphericity. The fit (curve 3) is complementary to the energy fit on Figure~\ref{Figure:Fit_AE}.}
    \label{Figure:Fit_R3}
\end{figure}

The limits obtained ($S^{\rm lim}_{avg} \simeq 1.7\times10^{-3}$~counts/(100 t day) at 90\% c.l.) are very low, e.g. $\sim 8\times10^4$ times lower than expected number of events from solar $pp-$ neutrino (135 counts/(100 t day)). The upper limits on the number of events with energy 5.5 MeV constrain the product of axion flux $\Phi_A$ and the interaction cross section  with electron, proton or carbon nucleus $\sigma_{A-e,p,C}$ via
\begin{equation}
S_{\rm events} = \Phi_{A}\sigma_{A-e,p,C}N_{e,p,C}T \leq S^{\rm lim} \label{Events},
\end{equation}
where  $N_{e,p,C}$ is the number of electrons, protons and carbon nuclei and $T$ is the measurement time, the product of these factors corresponds to the statistics of 579.3 kton$\times$day. 
The individual rate limits are:
\begin{eqnarray}
\Phi_{A}\sigma_{A-e} \leq 5.3\times 10^{-40} \rm{s^{-1}}\\  \label{SNU_e} \Phi_{A}\sigma_{A-p} \leq 3.4\times 10^{-39}
 \rm{s^{-1}}\\\label{SNU_p}
 \Phi_{A}\sigma_{A-C} \leq 4.5\times 10^{-39} \rm{s^{-1}}.\label{SNU_C}
\end{eqnarray}
The limits on the model-independent value $\Phi_{A}\sigma_{A}$ are very low, for comparison the standard solar neutrino unit is SNU = $10^{-36}\rm{s^{-1}} \rm{atom^{-1}}$ only and a capture rate of solar neutrinos by Ga-Ge radiochemical detectors is about 70 SNU.

\subsection{Limits on $|g_{A\gamma}\times g_{3AN}|$-coupling}
The number of $A\rightarrow 2\gamma$ decays in a volume V is:
\begin{equation}
N_{\gamma}=\Phi_{A}V\tau_f/\tau_{2\gamma},
\label{N_gamma}
\end{equation}
here $\Phi_{A}(m_A)$ is the flux of axions that have reached the detector, taking into account possible $A\rightarrow 2\gamma$-decay on the way from the Sun (Fig. \ref{Figure:Fluxes}) and $\tau_{f}$ is the time of flight of distance $L$ in the reference system associated with the axion:
\begin{equation}
\tau_{f}= (m_A/E_A)(L/\beta c)\label{tauf}.
\end{equation}
Here $\beta= \upsilon/c = p_{A}/E_{A}$ is the axion velocity in terms of the speed of light.

The number of events detected in the FV due to axion decays within the  Monte-Carlo volume (sphere with R=5~m) are:
\begin{equation}
S_{2\gamma} = N_{\gamma}T \varepsilon_{2\gamma} \label{Counts_2G} 
\end{equation}
where $N_{\gamma}$ is given by (\ref{N_gamma}), $T$ is live time measurements  
and $\varepsilon_{2\gamma}$ is the detection efficiency obtained by MC simulation (Table \ref{tab:Efficiency}). 
The expected value of $S_{2\gamma}$ has a complex dependence on $g_{A\gamma}$, $g_{3AN}$ and $m_A$ given by equations (\ref{tau_CM}-\ref{Flux}, \ref{N_gamma}-\ref{Counts_2G}).
In the assumption that $\beta \approx 1$ the number of $A\rightarrow2\gamma$ detected decays depends on $g_{3AN}^2$, $g_{A\gamma}^2$ and $m_A^4$ as:
\begin{equation}
S_{2\gamma}= 6.6\times 10^{4}g^2_{A\gamma} \times g_{3AN}^2\times m_A^4, \label{NG}
\end{equation}
where $g_{A\gamma}$ and $m_A$ are given in $\rm{GeV}^{-1}$  and eV units, respectively. 
The model-independent limit derived from the relation $S_{2\gamma} < S^{\rm lim}_{2\gamma}$   is  
\begin{equation}
 |g_{A\gamma}\times g_{3AN}|\times m_A^2 \leq 1.6\times 10^{-11} \rm{~eV~(90\%~c.l.)}. \label{AGMA}
\end{equation}

The relationship between the isovector constant $g_{3AN}$ and the axion mass $m_A$ allows to constraint the $g_{A\gamma}$ depending on $m_A$.
The three cases mentioned above were considered - the KSVZ-model with $E/N=0$ and the DFSZ-model for the angles $\tan\beta^*=140$ and $\tan\beta^*=0.25$ which are the boundary values of the allowed unknown angle $\beta^*$.

\begin{figure}
\includegraphics[width=9cm,height=10.5cm]{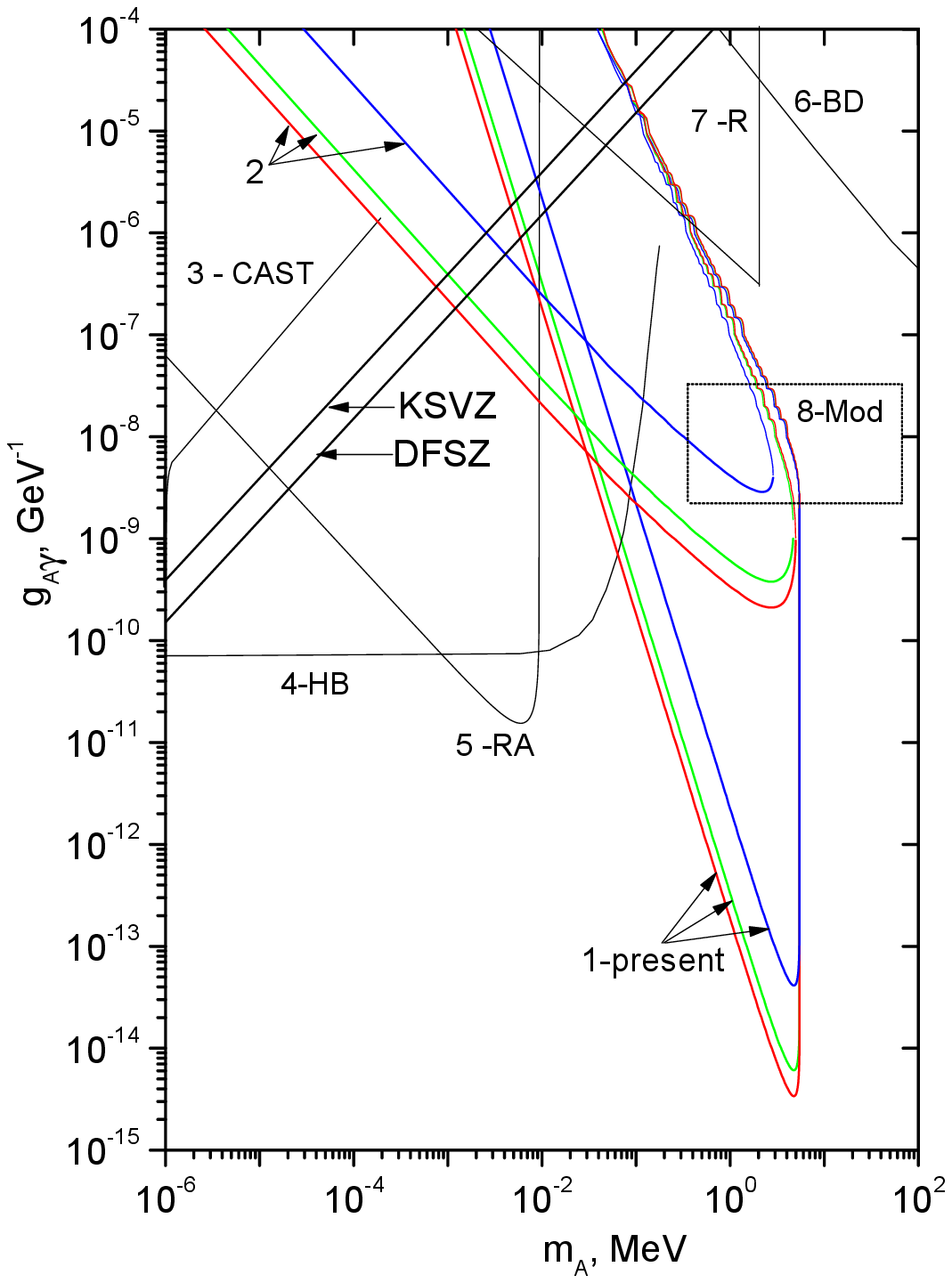}
\caption {The limits on $g_{A\gamma}$ obtained by 1 - present work ($A\rightarrow 2\gamma$-decay (from bottom to top: DFSZ ($\tan\beta^*=140$, red line),  KSVZ (green), DFSZ ($\tan\beta^*=0.25$, blue))), 2 - present work (PC, the same notations as for 1), areas of excluded values are located inside contour, 3- CAST \cite{Zio05,Ari09}, 4 - HB Stars \cite{Raf08}, 5 - resonant absorption \cite{Gav18}, 6 -  beam dump experiments \cite{Kon86,Bjo88}, 7-  reactor experiments \cite{Alt95,Cha07}, 8 - expectation region from heavy axion models \cite{Bere00,Bere01,Hal04}. Some data taken from \cite{Oha23}.
The relations between $g_{Ae}$ and $m_A$ for DFSZ- and KSVZ-models are shown also.}
\label{Figure:gag_limits}
\end{figure}

The excluded regions are inside three contours 1 in Fig.~\ref{Figure:gag_limits} (90 \% c.l.).
The curves from bottom to top correspond to DFSZ ($\tan\beta^*=140$), KSVZ and DFSZ ($\tan\beta^*=0.25$), respectively. 
For higher values of $g_{A\gamma}^2m_A^3$ axions decay before they reach the detector, while for lower $g_{A\gamma}^2m_A^3$ the probability of axion decay inside the Borexino volume is too low. The limits on $g_{A\gamma}$ obtained by other experiments are also shown.

The limits on $A\rightarrow2\gamma$ decay exclude a large new region of $g_{A\gamma}$ coupling constant ${(3\times10^{-15} - 10^{-8})~\rm{GeV}^{-1}}$ for the mass range $(0.02 - 5)$ MeV. 
The Borexino constrains are about 2-5 orders of magnitude stronger than those obtained by laboratory-based experiments using nuclear reactors and accelerators and partly overlap the  regions of heavy axion models \cite{Bere00,Bere01,Hal04}.

Taking into account the relation between $g_{eAN}$ and $m_A$ the constraint on $g_{A\gamma}$ and $m_A$ is given by
\begin{eqnarray}
 |g_{A\gamma}\times m_A^3| \leq 1.8\times 10^{-4} \rm{~eV^2~(\tan\beta^*=140)}\\
 |g_{A\gamma}\times m_A^3| \leq 3.2\times 10^{-4} \rm{~eV^2~(KSVZ)}\\
 |g_{A\gamma}\times m_A^3| \leq 2.2\times 10^{-3} \rm{~eV^2~(\tan\beta^*=0.25)}
   \label{AGMA1}
\end{eqnarray}
So, in the case of DFSZ-axion with $\tan\beta^*=140$ the coupling constant of 1~MeV axion is less than $g_{A\gamma}\leq1.8\times 10^{-13}\rm{~GeV^{-1}}$.

The figure \ref{Figure:gag_limits} shows the dependence of the $g_{A\gamma}$-constant on the mass $m_A$ calculated in accordance with expression (\ref{C_gamma}) for the values $E/N$ = 0 ($C_{A\gamma\gamma}$=-1.92) and $E/N$ = 8/3 ($C_{A\gamma\gamma}$=0.74) for the KSVZ- and DFSZ-models, respectively.

The number of expected events due to inverse Primakoff conversion is:
\begin{equation}
S_{PC} = \Phi_A\sigma_{PC}N_{C}T\varepsilon_{PC} \label{Counts_PC}
\end{equation}
where $\sigma_{PC}$ is the Primakoff conversion cross sections; $N_{C}$ is the number of  carbon nuclei in the 5~m radius sphere (Monte-Carlo (M-C) target), $T$ is live time of measurement and $\varepsilon_{PC}$ is the detection efficiency for 5.5~MeV $\gamma$'s (Fig.\ref{Figure:Response_Functions} and table \ref{tab:Efficiency}). 
The cross section $\sigma_{PC}$ is proportional to the $g^2_{A\gamma}$ and  the axion flux $\Phi_A$ is proportional to the  $g^2_{3AN}$ in accordance with (\ref{Primakoff}) and (\ref{FluxA}). 
The number of Primakoff conversions  $S_{PC}$ value depends on the product of the axion-photon and axion-nucleon coupling constants: $g^2_{A\gamma}\times g^2_{3AN}$. 
From the experimentally found relation $S_{PC}\leq S^{lim}_{PC}$ and for $(p_A/p_{\gamma})^3 \approx 1$  one obtain the model-independent upper limit on the product of constants:
\begin{equation}
 |g_{A\gamma}\times g_{3AN}|\leq 2.3\times 10^{-11}\rm{~GeV^{-1}} ~{\rm (90\%~c.l.).} \label{AAGPC}
\end{equation}

In the DFSZ- and KSVZ- axion models, the constraints on $g_{A\gamma}$ and $m_A$ are given by the relations:
\begin{eqnarray}
 |g_{A\gamma}\times m_A| \leq 3.2\times 10^{-13} \rm{~(\tan\beta^*=140)}\\
 |g_{A\gamma}\times m_A| \leq 5.7\times 10^{-13} \rm{~(KSVZ)}\\
 |g_{A\gamma}\times m_A| \leq 3.9\times 10^{-12} \rm{~(\tan\beta^*=0.25)}
   \label{AGMAPC}
\end{eqnarray}

For DFSZ axion with $m_A$=1 MeV and $\tan\beta^*=140$, the upper limit on $g_{A\gamma}$ corresponds to $g_{A\gamma}\leq3.2\times 10^{-10}\rm{~GeV^{-1}}$ that for $m_A$=1 MeV is significantly weaker than the $A\rightarrow 2\gamma$ limit. 
But for axion masses $m_A$ less than $\simeq$50 keV, the limits obtained from the Primakoff conversion become more stringent.
The regions of excluded values of $g_{A\gamma}$ and $m_A$ are shown in Fig.\ref{Figure:gag_limits} for three variants of axion models (three lines of contour 2).

The intersections of the KSVZ- and DFSZ-lines with the contours 1 and 2 in figure \ref{Figure:gag_limits} defines the region of axion masses excluded by this experiment: for KSVZ-axion the region (0.7-80)~keV and for DFSZ-axion $(\tan\beta^*=140)$ the region (1.2-100)~keV are excluded, respectively. 
The constraints on the coupling constant $g_{A\gamma}$ are the most stringent in comparison with the other experiments in axion mass range (0.6-1.0) MeV; if the decay $A\rightarrow e^+e^-$ is not taken into account, the range is expanded to 5 MeV.

\subsection{Limits on $|g_{Ae}\times g_{3AN}|$-coupling}

The number of Compton conversion events in the detector equals to:
\begin{equation}
S_{CC} = \Phi_{A}\sigma_{CC}N_{e}T\varepsilon_{CC} \label{Counts_CC}
\end{equation}
where $\sigma_{CC}$ is the Compton conversion cross sections, $\Phi_A$ is the axion flux on the Earth (Eq.~\ref{FluxA}), $N_{e}$ is the number of electrons in 5~m radius sphere  M-C target, the product $N_eT$ is the total statistics of the experiment and $\varepsilon_{CC}$ is the detection efficiency obtained with MC simulations (Fig.\ref{Figure:Response_Functions} and table \ref{tab:Efficiency}).

The $S_{CC}$ value depends on the product of the axion-electron and axion-nucleon coupling constants: $g^2_{Ae}\times (g_{3AN})^2$ because $\Phi_A$ is proportional to the $(g_{3AN})^2$ and $\sigma_{CC}$ is proportional to the $g^2_{Ae}$ according to expressions (\ref{FluxA}) and (\ref{CC_CS}). 
The dependence of $|g_{Ae}\times g_{3AN}|$ on $m_A$ arises from the kinematic factor in equations for $\Phi_A(m_A)$~(\ref{FluxA}) and $\sigma_{CC}(m_A)$~(\ref{CC_CS}). 

The experimental $S_{CC}^{\rm lim}$ can be used to constrain $g_{Ae}\times g_{3AN}$ and $m_A$. 
At approximate equality of the momenta of the axion and the $\gamma$-quantum ($(p_A/p_{\gamma})^3 \simeq 1$ for $m_A\leq 1$ MeV) the limit is:
\begin{equation}
|g_{Ae}\times g_{3AN}| \leq 1.9\times10^{-13}   ~\rm{(90\% ~c.l.).} \label{gae_gan_limit}
\end{equation}
 This constraint is completely model-independent and valid for any pseudoscalar particle.
 The previous limit is improved by a factor of $\sim 3$ \cite{Bel12a}, but one must take into account the complicating fact that the expected number of axion events is proportional to the square of the product $g_{Ae}$ and $g_{3AN}$ constants: $S_{CC} \sim |g_{Ae}\times g_{3AN}|^2$.
 It's important to stress that the limits were obtained on the
assumption that axions escape from the Sun and reach the Earth, which implies $g_{Ae} < 10^{-6}$ for $m_A < 2m_e$ and $g_{Ae} < (10^{-11}-10^{-12})$ if
$m_A > 2m_e$ (\cite{Der10,Bel12a}).

\begin{figure}
\includegraphics[width=9cm,height=10.5cm]{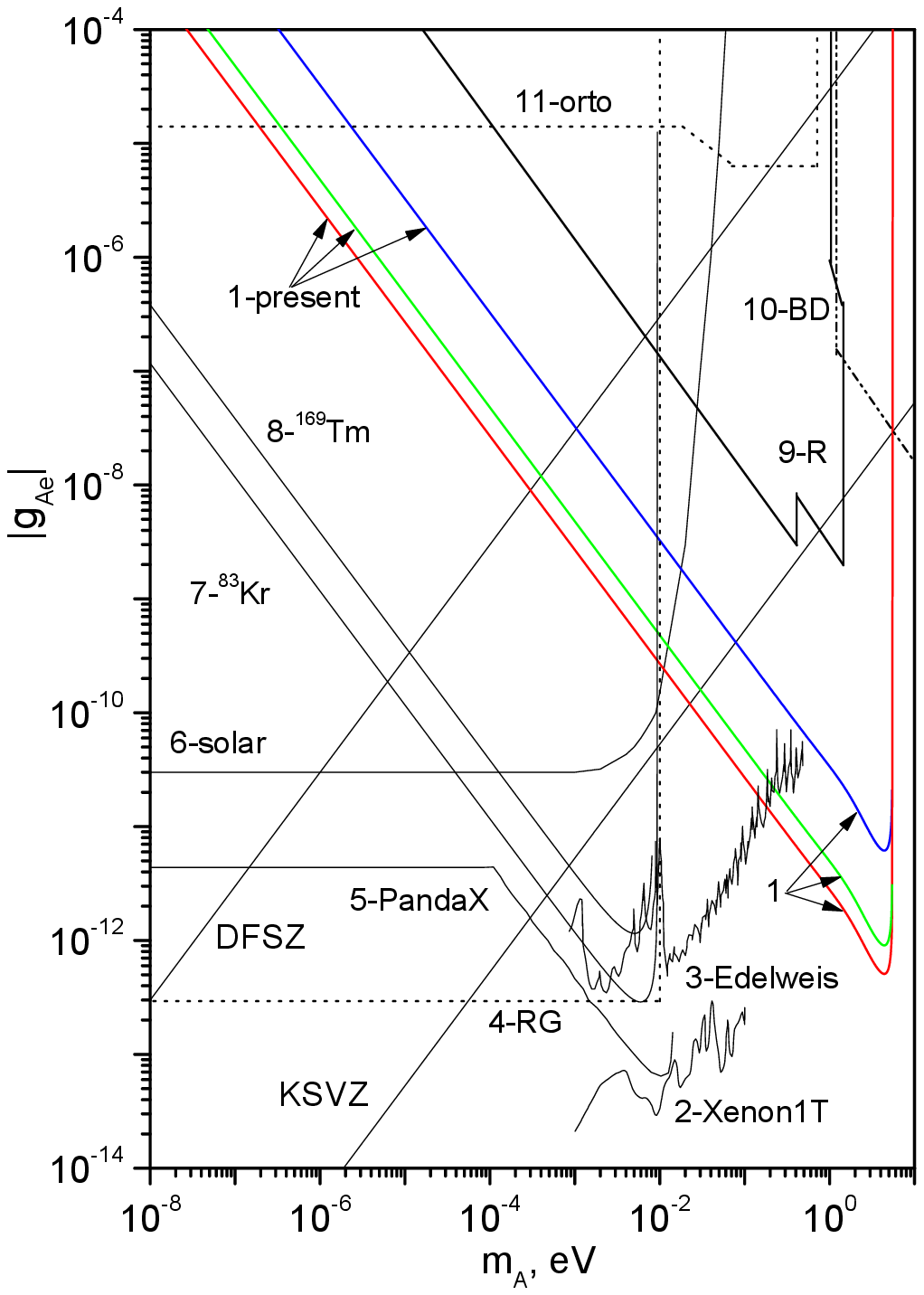}
\caption {The limits on the $g_{Ae}$ obtained by 1 - present work (Compton process $A+e\rightarrow e+\gamma$ (from bottom to top: DFSZ ($\tan\beta^*=140$, red line),  KSVZ (green), DFSZ ($\tan\beta^*=0.25$, blue)), 2 - XENON1T \cite{Apr18} (relic axions),  3 - Edelweiss \cite{Arm18} (relic axions), 4 - red giants \cite{Raf08} , 5- PandaX-II \cite{Fu17} (relic and solar axions), 6 - solar axion luminosity \cite{Gon09}, 7 - ${\rm{^{83}Kr}}$ \cite{Gav22} (resonant absorption of solar axions) , 8 - ${\rm{^{169}Tm}}$ \cite{Der09,Abd20} (resonant absorption of solar axions) , 9 - reactor \cite{Alt95,Cha07} and solar experiments \cite{Bel08,Der10}, 10 - beam dump experiments \cite{Kon86,Bjo88}, 11 - ortho-positronium decay \cite{Asa91}. 
Some data taken from \cite{Oha23}.
The relations between $g_{Ae}$ and $m_A$ for DFSZ- and KSVZ-models are shown also.} \label{Figure:gae_limits}
\end{figure}

Using the relationship between  $g_{3AN}$  and $m_A$ in KSVZ- and DFSZ-models one can obtain a constraint on the $g_{Ae}$ constant, depending on the axion mass (Fig. \ref{Figure:gae_limits}, three lines marked by 1). 
For $(p_A/p_{\gamma})^3 \approx 1$ the limit on $g_{Ae}$ and $m_A$ is at 90\% ~c.l.:
\begin{eqnarray}
|g_{Ae}\times m_A| \leq 2.7\times10^{-6}~\rm{eV} ~(\tan\beta^*=140)\\
|g_{Ae}\times m_A| \leq 4.8\times10^{-6}~\rm{eV} ~\rm{(KSVZ)}\\
|g_{Ae}\times m_A| \leq 3.3\times10^{-5}~\rm{eV} ~(\tan\beta^*=0.25)
\label{gae_ma_limit}
\end{eqnarray}
where $m_A$ is given in eV units. 
For DFSZ-axion with $m_A$=1 MeV and $\tan\beta^* = 140$, the upper limit corresponds to $g_{Ae}\leq 2.7 \times 10^{-12}$. 
The upper limit on KSVZ-axion mass is $m_A\leq20$~keV and for DFSZ-axion $(\tan\beta^*=140)$  - $m_A\leq300$~eV.

Figure \ref{Figure:gae_limits} shows additionally the constraints on $g_{Ae}$ that were obtained in experiments searching for axio-electric effect of relic axions \cite{Apr18,Arm18}, 14.4 keV solar axions \cite{Fu17}, resonant absorption of solar axions \cite{Gav22,Der09,Abd20}. Also results of reactor  \cite{Alt95,Cha07} and beam-dump \cite{Kon86,Bjo88} experiments as well as constraints from astrophysical arguments \cite{Raf08,Gon09} and ortho-positronium decay \cite{Asa91} are shown.

\section{Conclusions}
A search for signals from solar axions emitted in the $p(d,\rm{^3He})A$ $\rm{(5.5~MeV)}$ reaction has been performed with the complete dataset of the Borexino detector. 
The flux of such axions depends on the isovector coupling constant of the axion with nucleons $g_{3AN}$. 
The decay of axions into two photons ${A\rightarrow2\gamma}$, inverse Primakoff conversion on nuclei ${A+\rm{^{12,13}C}\rightarrow \rm{^{12,13}C}+\gamma}$ and the Compton process ${A+e\rightarrow e+\gamma}$  were studied. 
The cross section of the first two reactions depends on the coupling constant of the axion with photons $g_{A\gamma}$, the latter on the coupling constant of the axion with electrons $g_{Ae}$.

The signature of all these reactions is a $5.5$~MeV peak in the energy spectrum of the Borexino detector. No statistically significant indications of axion interactions were detected. 
New model-independent upper limits on the product of the axion coupling constants to photons, electrons and nucleons were obtained:
\begin{equation}
|g_{A\gamma}\times g_{3AN}|\times m_A^2 \leq 1.6\times 10^{-11} \rm{eV},
\end{equation}
\begin{equation}
 |g_{A\gamma}\times g_{3AN}|\leq2.3\times 10^{-11} \rm{GeV^{-1}},~ \rm{and}
\end{equation}
\begin{equation}
 |g_{Ae}\times g_{3AN}|\leq 1.9\times10^{-13},
\end{equation}
all at 90\% c.~l.
Compared to previous work~\cite{Bel12a}, the limits are improved by a factor of (2 – 3), that corresponds to roughly an order-of-magnitude increase in sensitivity to the count rates.

From these model-independent limits, we derive constraints on the products $|g_{A\gamma}\times m_A|$ and $|g_{Ae}\times m_A|$ for the KSVZ-axion model and the DFSZ-axion model for different values of $\tan\beta^*$.
The new Borexino results exclude large regions of axion-photon $g_{A\gamma}$, axion-electron $g_{Ae}$ coupling constants and axion masses $m_A$, and allow us to constrain the values of the axion masses in specific variants of the KSVZ- and DFSZ-axion models.

\section{Acknowledgments}
The Borexino program is made possible by funding from Istituto Nazionale di Fisica Nucleare (INFN) (Italy), National Science Foundation (NSF) (USA), Russian Science Foundation (RSF) (Grant No. 24-12-00046), Deutsche Forschungsgemeinschaft (DFG), Cluster of Excellence PRISMA+ (Project ID 39083149), and recruitment initiative of Helmholtz-Gemeinschaft (HGF) (Germany), and Narodowe Centrum Nauki (NCN) (Grant No. UMO-2013/10/E/ST2/00180) (Poland). 
We acknowledge the generous hospitality and support of the Laboratori Nazionali del Gran Sasso (Italy).

\end{document}